\documentclass[twoside,psfig]{article}
\usepackage{fleqn,espcrc2,epsfig}

\usepackage{graphicx}
\usepackage[figuresright]{rotating}

\def\kk{{\bf k_1}}
\def\kg{{\bf k_2}}
\def\qq{{\bf q}}
\newcommand{\AmS}{{\protect\the\textfont2
  A\kern-.1667em\lower.5ex\hbox{M}\kern-.125emS}}

\title{Finite Gluons Ladders and Hadronic Collisions \thanks{This work was
partially supported by CNPq, Brazil.}}

\author{M. B. Gay Ducati  and  M. V. T.
Machado  \address[IFUFRGS]{Instituto de F\'{\i}sica, Univ.
Federal do Rio Grande do Sul.\\  Caixa Postal 15051, 91501-970 Porto Alegre,
RS, BRAZIL.} }
       
\begin{document}

\begin{abstract}
A truncated BFKL series is studied and applied to  hadronic
processes. The $\sigma^{pp(p\bar{p})}_{tot}$ are described
with  good agreement with data and in a way consistent with the unitarity
bound. The elastic  scattering amplitude is calculated at $t \neq 0$,
introducing two distinct ans\"atze for the proton impact factor. The
$d\sigma^{el}/dt$ is obtained at small $t$
approximation and compared with the data.

\vspace{1pc}
\end{abstract}

% typeset front matter (including abstract)
\maketitle

\section{Introduction}
The understanding of  the  BFKL Pomeron has been  demanding a considerable theoretical effort. Its behavior 
in perturbative QCD is determined by generating the
integral equation \cite{bfkl}. That procedure consists of summing the leading logarithms on
energy, $\ln(s)$, order by order from perturbation theory and the main result
is that the $\sigma_{tot}$ for the $I\!\!P$-exchange process is a power of
$s$. 

A priori, BFKL is itself asymptotic and we may ask if at finite
energies, i.e. non asymptotic regime, summing a finite number of terms
from the  BFKL series   could describe the existent data.  For that we should use a truncated BFKL series, performing a finite sum of
gluon ladders (bearing in mind reggeized gluons and considering effective
vertices). The question that remains is how many orders should one  take into
account. The lowest order two gluons exchange calculation leads to a
$\sigma_{tot}$  constant on $s$ and the next contribution to the sum is the
one rung gluon ladder, which provides a logarithmic growth. In order to avoid
unitarity violation and  by simplicity we truncate our summation at this
order.  As a result a successfull fit to the $\sigma^{pp(p\bar{p})}_{tot}$ 
with these two contributions is obtained and presented in Sec. 2. These
results motivate to check the non-forward amplitude in order to obtain the
prediction for the $d\sigma^{el}/dt$, which gives the behavior on the momentum
transfer $t$.

In the BFKL framework such analysis is dependent of the proton impact 
factor (IF) input, which introduces certain  uncertainty due to the
presence of non-perturbative content. The IFs
determine the coupling of the Pomeron to the color singlet hadrons and play a
crucial role in the calculation of the non-forward amplitude. We calculate
the $pp(p\bar{p})$ elastic amplitude at $t \neq 0$ taking into
account two distict ans\"atze to the proton IF: the Dirac form
factor, as proposed recently by Balitsky and Kuchina
\cite{balitsky} and the usual non-perturbative ansatz \cite{askew}. The main
resulting features are discussed in Sec. 3, although considering that a more
realistic ansatz to the proton IF is still to be found. In addition the
$d\sigma^{el}/dt$ is calculated in the small $t$
approximation and compared with the experimental data. In the last section we present our conclusions. 

\section{Truncated BFKL Series}
In the leading logarithm approximation (LLA), the Pomeron is obtained
considering the color singlet ladder diagrams whose vertical lines are
reggeized gluons coupled to the rungs through the effective vertices. The
correspondent amplitude is purely imaginary and the coupling constant
$\alpha_s$ is considered frozen in some transverse momentum scale \cite{bfkl}.
For the elastic  scattering of a hadron, the Mellin transform of the
scattering amplitude is  given by \cite{forshaw}:

\begin{eqnarray} 
{\cal A}(\omega, t) &=& \frac{{\cal G}}{(2 \pi)^2} \int d^2 \kk \, d^2 \kg
\,\frac{\Phi(\kk)\Phi(\kg)}{\kg^2(\kk-\qq)^2}\, \nonumber
\\
& &\times f(\omega,\kk,\kg,\qq)  \,,   
\end{eqnarray}
\noindent where the ${\cal G}$ is the color factor for the color
singlet exchange, $\kk$ and $\kg$ are the transverse momenta of the exchanged
gluons in the $t$-channel and $\qq$ is the momentum transfer, with $\qq^2=-t$.

The function $f(\omega,\kk,\kg ,\qq)$  is the Mellin transform of the BFKL kernel 
$F(s,\kk,\kg,\qq)$, which states the dynamics of the process and is
obtained in perturbative QCD. The main properties of the LO kernel are well
known \cite{bfkl} and the results arising from the NLO calculations have
yielded intense debate in the literature recently \cite{NLO}.

In the case of $pp$($p\bar{p}$) scattering, the factor $\Phi({\bf k})$ is the
proton IF, which in the absence of a perturbative scale has a
non-perturbative feature and furnishes the $I\!\!P$-proton coupling. This
factor turns the amplitude infrared safe when the colliding particles are
colorless. In the leading order of  perturbation theory we have 
\begin{eqnarray}
f_1(\omega,\kk,\kg,\qq) = \frac{1}{\omega}\,\delta^2(\kk-\kg)\,,
\end{eqnarray} 
\noindent and in the next order
 \begin{eqnarray} f_2(\omega,\kk,\kg,\qq)=
-\frac{\bar{\alpha}_s}{2 \pi}\,\frac{1}{\omega^2} \left[
\frac{\qq^2}{\kk^2(\kg-\qq)^2} \right. \nonumber \\ 
-
\left.
\frac{1}{2}\frac{1}{(\kk-\kg)^2}\left(1+\frac{\kg^2(\kk-\qq)^2}{\kk^2(\kg-\qq)^2} \right)  \right]  \,. \end{eqnarray}

\noindent Here, $\bar{\alpha}_s=N_c\alpha_s/\pi$, where $N_c$ is the
color number and $\alpha_s$ is the strong coupling constant.  In order to perform a reliable calculation the
convenient proton IF should be introduced. This is not an easy
task, namely these hadronic processes are soft and there is no hard scale
allowing to use perturbation theory.  In fact, we should know in details the
parton wavefunction in the hadron to calculate the IFs properly.
Since this is not available, several models are proposed in order to calculate
them. 

From the Optical Theorem, the lowest
order contribution to the $\sigma_{tot}$ is a constant term, and the next
order term  is a logarithm of the energy,  scaled by a typical gluon
transverse momentum of the process (arbitrary). When considering $t=0$ there
is no need to deal with both a specific form for the IF and the transverse
momentum integration. This allows to consider $s$-independent factors in each
term as free parameters and to obtain them from data. The correct description at low energy requires the reggeon
contribution, which is parameterized from Regge theory. Our expression to the
total cross section is then,

\begin{eqnarray} 
\sigma_{tot}^{pp(p\bar{p})}= C_{R}(\frac{s}{s_0})^{\alpha_{R}(0)-1} +
C_{Born} + C_{HO}\ln (\frac{s}{s_0})\,. \nonumber
\end{eqnarray}
Hence we fix the constants $C_{Born}$
and $C_{HO}$ from data on $p\bar{p}$, imposing the same contribution for
both $pp$ and $p\bar{p}$. This procedure is reasonable due to
the higher energies reached on $p\bar{p}$ collision, where the Pomeron
dominates. On the other hand, $pp$ data are predominantly at low energy, which
is not strongly sensitive to the Pomeron model, thus dominated by the 
reggeonic contribution. A successful description
of data  is obtained for the whole range of energy. The result is shown in the
Fig. (1). The parameters  and a more detailed discussion can be
found in Ref. \cite{finite}.

\begin{figure}[h]
\centerline{\psfig{file=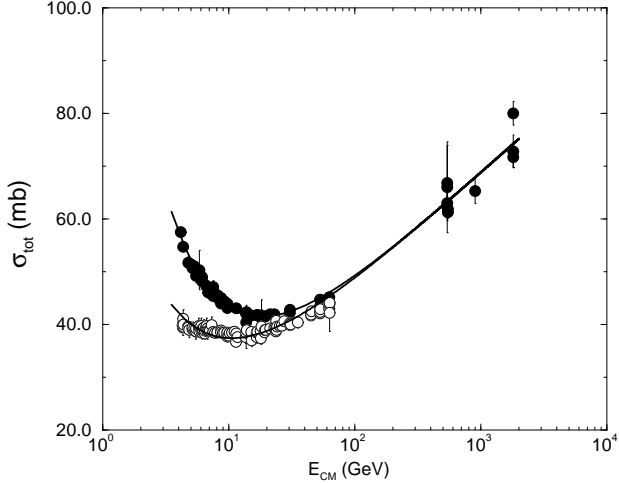,width=90mm}} 
\caption{Result of the $pp(p\bar{p})$ total cross sections \cite{data}.
The errors are added into quadrature.} 
\end{figure}

The hypothesis of considering two orders from the BFKL series
($\sigma_{tot}\sim \ln (s)$) is phenomenologically corroborated by the well
known dispersion relation fit \cite{matthie}. An important additional
advantage is that  the total cross section obtained is consistent with the
unitarity constraint, avoiding unitarization procedures. As a final remark, at
the LHC energy ($\sqrt{s}=14$ TeV) the extrapolation of our results will give
$\sigma_{tot}=93.22\,\,mb$.

\section{The non forward scattering amplitude}

In order to calculate the elastic amplitude at $t \neq 0$,
information about the coupling between the proton and the $t$-channel
reggeized gluons in the ladder is required. Namely, introducing a
reliable proton IF that has to be 
modeled since it cannot be calculated from first principles due to the
unknowledge on the wavefunction of the hadronic constituent partons. Here are
analized two distinct models for the IF:

\subsection{Dirac form factor:}
Balitsky and Kuchina proposed recently \cite{balitsky} that at large momentum transfer the coupling of the
BFKL Pomeron to the nucleon is essentially equal to the Dirac form factor of
the nucleon. Their basic idea is that in the lowest order in perturbation
theory there is no difference between the diagrams for the nucleon IF and
similar diagrams with two gluons replaced by two photons, in such a way that
the amplitudes  can be calculated without any model assumption.

This IF, $\Phi_p({\bf k},\qq)$,  is  decoupled in the transverse
momentum integration and presents an explicit dependence on $t$, being
similar to the usual Pomeron-proton coupling used in Regge phenomenology. The
expression is

\begin{eqnarray}
F_1^{p+n}(t) =
\frac{1}{1+\left(\frac{|t|}{0.71\,GeV^2}\right)^2}\frac{4m_p^2+0.88|t|}
{4m_p^2+|t|}\,. \end{eqnarray}

The choice for this proton IF is useful when one analyzes near
forward observables, for instance the elastic differential cross section. However
it does not play the role of a regulator of infrared divergences at
$pp(p\bar{p})$  process  because clearly it does not vanish when
the gluon transverse momenta goes to zero. In electron-proton process the
situation is different  since the photon impact factor supplies that condition
\cite{balitsky}. 

Then the next step is to perform the gluon transverse momenta integrations. In
fact, such integrals are infrared divergent and should be regularized. An
usual way out is to introduce an  infrared cut-off $\lambda^2$, temporally defining a small gluon mass, avoiding
problems at the infrared region. This procedure is quite similar as to take
into account a non-perturbative massive gluon propagator (see i.e. Ref.
\cite{halzen}).  

The lowest order (order $\alpha_s^2$) contribution, using Eqs. (1-2), gives the
following result: 
\begin{eqnarray} 
{\cal A}^{(1)}(s,t)= \frac{{\cal
G}^{\prime}}{(2\pi)^4}\,s\, [F_1^{p+n}(t)]^2 \frac{\pi}{(|t|-\lambda^2)}\,\ln
\left(\frac{\lambda^2}{|t|}\right) \,. \nonumber \end{eqnarray}

The one rung gluon ladder has two components (order $\alpha_s^3$), given by the following expression:
\begin{eqnarray}
{\cal A}^{(2)}(s,t)= \frac{{\cal G}^{\prime}}{(2\pi)^4}\,[F_1^{p+n}(t)]^2
\,s\, \ln\left(\frac{s}{{\bf k^2}}\right)\,(I_1 + I_2)\,,
\nonumber \end{eqnarray} with $I_1$ corresponding to the one rung gluon ladder
and $I_2$ correspondent to the three gluons exchange graphs, whose order is
also $\ln(s/{\bf k^2})$. Such structure is due to the fact that in the color
singlet calculation there is no cancellation between graphs and one can not 
obtain an expression for the two-loop level which is proportional to the one
loop amplitude \cite{forshaw}. We define $I_2$ through symmetry on the
integration variables $\kk$ and $\kg$ (see Eqs. (1,3)) and the factor ${\cal
G}^{\prime}$ collects the correspondent color factors and the remaining
constants. The explicit calculation of those integrals, yields

\begin{eqnarray}
I_1=  -\pi^2 \frac{|t|}{(|t|-\lambda^2)^2}\,\ln^2
\left(\frac{\lambda^2}{|t|} \right)\,, \nonumber \end{eqnarray}

\begin{eqnarray}
I_2=
\frac{1}{2}\frac{\pi^2\,\ln(\lambda^2)}{(|t|-\lambda^2)}
\ln\left(\frac{\lambda^2}{|t|}\right)\left(
1-\frac{\ln(|t|)}{\ln(\lambda^2)}\right)\,. \nonumber \end{eqnarray}

Some comments about the amplitude above are in order. The scale of the factor
$\lambda^2$ should be at non-perturbative regime, i.e. $< 1$ GeV$^2$.
An interesting aspect is the behavior of the amplitude at the forward limit
$t=0$, where it becames very large. This limit is a well known property of
perturbative QCD calculations and there are several reasons to believe that
the point $t=0$ plays a very special role, such that perturbation theory
may even not be applicable. For the full BFKL series in the forward region
there is still the diffusion on tranverse momenta, i.e. on $\ln {\bf k}^2$,
which extends into both the ultraviolet and the infrared regions
\cite{forshaw}. Nevertheless, the momentum scale $t$ supplies  the control
condition.

However, we will suppose that a smooth transition from a finite $t$ down to
$t=0$ is possible and that the finite BFKL series gives the correct behavior on
energy for the forward observables. Later we make use of this hypothesis to
obtain the logarithmic slope $B(s)$ and the differential elastic cross section.

\subsection{Usual non-perturbative ansatz:}

Using quite general properties of the IFs, namely they vanish as
transverse momenta go to zero, one can guess their behavior which is
determined by the large scale nucleon dynamics. Regardless its exact shape, in
general the proton IF takes the form \cite{askew}: 

\begin{eqnarray}
\Phi_p({\bf k})= \frac{{\bf k}^2}{{\bf k}^2 + \mu^2}\,,
\end{eqnarray}
where $\mu^2$ is a scale which is typical of the non-perturbative dynamics.
As a consequence of this choice the momentum transfer behavior is completely
determined by the kernel. The amplitude now reads: \begin{eqnarray} {\cal
A}(s,t)= A^{\prime}\,\left[ \frac{1}{(|t|-\mu^2)} 
+ \frac{|t|} {(|t|-\mu^2)^2} \ln \left(\frac{\mu^2}{|t|}\right)\right]  
\nonumber \\ + \, A^{\prime}\,\ln\left(\frac{s}{{\bf k}^2}\right)\,\left[ 
\frac{\ln (\mu^2)}{(|t|- \mu^2)} + \frac{\ln(\mu^2)|t|}{(|t|-\mu^2)^2}\ln
\left(\frac{\mu^2}{|t|}\right)\right] \,.\nonumber 
 \end{eqnarray} 
\noindent Here we use the definition $A^{\prime}=\frac{{\cal
G}}{(2\pi)^4}\,s\,\pi$.

 We observe
again a divergent behavior at $t=0$, nevertheless we claim that the forward
amplitude is finite in this point and the dependence on energy is correctly
described. Despite obtaining an analytic expression to the elastic scattering
amplitude, a direct comparison with the experimental data is known not to be
reliable. In fact, data on differential cross section at low $t$ are
parameterized in the form $d\sigma/dt=A\,e^{B\,t}$, where $B$ is the forward
slope \cite{matthie}. Therefore, we can obtain an expression for the
differential cross section at small $t$, using our previous results. 

The usual relation to
describe the cross section is:
 \begin{eqnarray} \frac{d\sigma^{el}}{dt} & = & 
\frac{d\sigma}{dt}|_{t=0}\,e^{B(s,\,t=0)\,t}=\frac{\sigma_{tot}^2}{16
\pi}\,e^{B_{el}(s)\,t}\,,\\ B(s) &  = &
\frac{d}{dt}\left[\log\,\frac{d\sigma}{dt}\right]_{t=0}\,. \end{eqnarray} 

In the
Regge framework the slope is obtained from the powerlike behavior of the
scattering amplitude, dependent of the effective slope of the Pomeron
trajectory $\alpha^{\prime}_{P}$, namely $B_{el}^{Regge}(s)= 4 b_0 +
2 \alpha^{\prime}_{P}\,\ln(s)$. The parameter $b_0$ comes from the slope of
the $p$-$p$-$I\!\!P$ vertex. In our case we should obtain the slope from the
non forward elastic scattering amplitudes ${\cal A}^{Ladder}(s,t)$ obtained
above. For the amplitude obtained employing the Balitsky and Kuchina
impact factor it results the following slope  \begin{eqnarray}
B(s)=\frac{4}{F_1^{p+n}(t)}\,\frac{d F_1^{p+n}(t)}{dt} + \frac{2}{{\cal
A}(s,t)}\frac{d\,{\cal A}(s,t)}{dt}\,\,|_{t=0},\nonumber \end{eqnarray}
where the first term  does not contribute effectively at $t=0$ and we are left
only with the second term. From simple inspection of the amplitude obtained
with the usual impact factor (see Eq. (5)) we also verify that one gets a
similar expression to the correspondent slope. 

Considering the  specific form for the $t$-derivative of the amplitudes, their
asymptotic values at $t=0$ depend only on the energy. In fact, they take the
form $d{\cal A}/dt=R_1\,s+ R_2\,s \ln(s/s_0)$, where $R_1$ and $R_2$ are
$s$-independent parameters. For our case, the amplitude is purely imaginary,
then $|{\cal A}(s,t=0) |= s\,\sigma_{tot}$ and $d\sigma/dt
\,|_{t=0}=\sigma_{tot}^2/16\pi$. Putting all together, the corresponding slope
is  \begin{eqnarray}
B(s) & = & \frac{2}{\sigma_{tot}}\left[R_1 + R_2\,\ln(s/s_0)\right]\,\,,
\end{eqnarray} 
\noindent and the elastic differential cross section is given by Eq. (6),
where again $s_0=1\,GeV^2$.

In order to obtain the parameters
$R_1$ and $R_2$, we use the slope experimental values for both low (CERN-ISR)
and high energy (CERN-SPS,Tevatron) points from $p\bar{p}$ reaction
($23\,\,$GeV$<\sqrt{s}<\,1800$ GeV) \cite{data}. Having the slope obtained from data,
the elastic differential cross section is straighforwardly determined and a
successful comparison with its experimental measurements
at $\sqrt{s}=1800$ GeV is shown in the  Fig. (2).
\begin{figure}[h]
\psfig{file=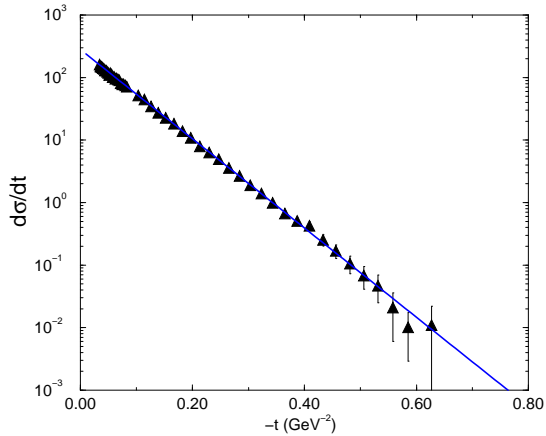,width=80mm}
\caption{The result for the elastic differential cross section at $\sqrt{s}=
1800$ GeV \cite{data}.} 
\end{figure}
In summary, we study the contribution of a  truncated BFKL series to
the hadronic process, specifically the $pp(p\bar{p})$ collisions, considering
two orders in perturbation theory corresponding to the bare two gluons exchange
and the one rung gluon ladder. Despite the restrictions imposed by the use of a
perturbative approach for soft observables, a good description of the total
cross sections was obtained motivating an analysis of the elastic differential
cross section. Although the QCD perturbation theory is in
principle not reliable at the forward direction ($t=0$), nevertheless we
suppose that perturbation theory gives the behavior on energy even in this
region.  The next step is to consider $t$ different from zero, where the
momentum transfer furnishes a scale to perform suitable calculations.  In
order to proceed this, we calculate the non forward amplitude introducing two
distinct ans\"atze for the proton impact factor, namely a factorizable
$t$-dependent proposed recently by Balitsky and Kuchina and the usual
non-perturbative impact factor. In order to describe data we used  a small
momentum transfer approximation and obtained an expression to the elastic slope
$B_{el}(s)$ and  the correspondent parameters. The elastic
differential cross section is obtained straightforwardly, describing with good
agreement the experimental data at both low and high energy values.

\section*{Acknowledgements}
MBDG thanks the Organizers of this lively meeting and the useful discussions
with P. Landshoff, A. Kaidalov, E. Levin, V. Petrov and M. Ryskin.

\end{document}